\documentclass[english,twocolumn,pra,eqsecnum, showpacs]{revtex4}
\usepackage[T1]{fontenc}
\usepackage[latin1]{inputenc}
\usepackage{amsmath}
\usepackage{graphicx}
\usepackage{amssymb}

\makeatletter
\makeatletter
\usepackage{epsf}

\makeatother

\usepackage{babel}
\makeatother
\begin{document}

\title{Mean field thermodynamics of a spin-polarized spherically trapped
Fermi gas at unitarity}

\author{Xia-Ji Liu$^{1}$, Hui Hu$^{1,2}$, and Peter D. Drummond${^{1}}$}

\affiliation{$^{1}$\ ARC Centre of Excellence for Quantum-Atom Optics, School
of Physical Sciences, University of Queensland, Brisbane, Queensland
4072, Australia \\
 $^{2}$\ Department of Physics, Renmin University of China, Beijing
100872, China}

\date{\today{}}

\begin{abstract}
We calculate the mean-field thermodynamics of a spherically trapped
Fermi gas with unequal spin populations in the unitarity limit, comparing
results from the Bogoliubov-de Gennes equations and the local density
approximation. We follow the usual mean-field decoupling in deriving
the Bogoliubov-de Gennes equations and set up an efficient and accurate
method for solving these equations. In the local density approximation
we consider locally homogeneous solutions, with a slowly varying order
parameter. With a large particle number these two approximation schemes
give rise to essentially the same results for various thermodynamic
quantities, including the density profiles. This excellent agreement
strongly indicates that the small oscillation of order parameters
near the edge of trap, sometimes interpreted as spatially inhomogeneous
Fulde-Ferrell-Larkin-Ovchinnikov states in previous studies of Bogoliubov-de
Gennes equations, is a finite size effect. We find that a bimodal
structure emerges in the density profile of the minority spin state
at finite temperature, as observed in experiments. The superfluid
transition temperature as a function of the population imbalance is
determined, and is shown to be consistent with recent experimental
measurements. The temperature dependence of the equation of state
is discussed. 
\end{abstract}

\pacs{03.75.Hh, 03.75.Ss, 05.30.Fk}

\maketitle

\section{Introduction}

There has been considerable recent experimental progress in creating
strongly interacting ultra-cold atomic Fermi gases. A primary tool
for the manipulation of these systems is the use of Feshbach resonances,
through which the magnitude and sign of the inter-atomic interaction
can be tuned arbitrarily by an external magnetic field. For a two-component
(\textit{i.e.}, spin 1/2) Fermi gas with equal spin populations, it
has been expected for some time that the system will undergo a smooth
crossover from Bardeen-Cooper-Schrieffer (BCS) superfluidity to a
Bose-Einstein condensate (BEC) of tightly bound pairs. This scenario
has now been confirmed unambiguously by some recent measurements on
both dynamical and thermodynamical properties \cite{jila,mit2004,duke2004,ins2004a,hui2004,ins2004b,duke2005,mit2005}.

Since the population in each spin state can also be adjusted with
high accuracy \cite{mit2006a,mit2006b,mit2006c,rice}, a subtle question
of particular interest is the ground state of a spin-polarized Fermi
gas with different particle numbers in the spin up and down states.
As conventional BCS pairing requires an equal number of atoms for
each spin component, exotic forms of pairing are necessary in order
to accommodate a finite spin population imbalance. There are several
scenarios suggested in the weakly coupling BCS limit for a uniform
gas, including the spatially modulated Fulde-Ferrell-Larkin-Ovchinnikov
(FFLO) state \cite{fflo}, the breached pairing \cite{bp} or Sarma
superfluidity \cite{sarma,wu,pao}, and phase separation \cite{bedaque}.
In the strong coupling, BCS-BEC crossover regime, a variety of mean-field
phase diagrams have been also proposed \cite{son,yang,sheehy,hui2006,xiaji1,bulgac1,mannarelli,chien1,parish}.
However, no clear consensus on the {\em true} ground state of spin-polarized
fermionic superfluidity has been reached as yet \cite{casalbuoni,dfs,carlson,he,caldas,ho2006,gu,iskin,yip1}.

Recent investigations \cite{mit2006a,mit2006b,mit2006c,rice} on atomic
$^{6}$Li gases with tunable population imbalance open up intriguing
possibilities for solving this long-standing problem. These experimental
observations have attracted intense theoretical interest \cite{chevy,yi,silva,haquea,yip2,imambekov,martikainen,chien2,bulgac2,gubbels,mizushima,castorina,kinnunen,machida,jensen}.
We note that there is no firm experimental evidence for the various
non-standard superfluid states mentioned earlier, which involve homogeneous
spin-polarized environments. Various interesting phenomena have been
demonstrated experimentally, in optical traps of different shapes
and sizes:

\begin{description}
\item [{{{(A)}}}] A shell structure is observed in the density profiles
by Zwierlein \textit{et al} \cite{mit2006b}, with a {\em bimodal}
distribution at finite temperature, suggesting an interior core of
a BCS superfluid phase with an outer shell of the normal component.
As a result, the thermal wing may provide a direct route to thermometry,
in an environment where temperature is often difficult to calibrate
reliably. 
\item [{{{(B)}}}] With increasing spin-polarization, the gas shows
a quantum phase transition from the superfluid to normal state. Close
to the broad Feshbach resonance of $^{6}$Li at $B\simeq833$ G, a
critical polarization $P_{c}=0.70(3)$ has been determined at low
temperatures. Here, the relative polarization is $P=(N_{\uparrow}-N_{\downarrow})/(N_{\uparrow}+N_{\downarrow})$
where $N_{\sigma}$ is the number of spin up or down atoms. The value
of $P_{c}$ decreases with increased temperature. 
\item [{{{(C)}}}] A similar shell structure is also identified by Partridge
\textit{et al} \cite{rice}\textit{,} in an experimental trap configuration
with a very large aspect ratio. However, this bimodal structure disappears
below a threshold polarization of $P_{*}\sim$ $0.10$. 
\end{description}

To make quantitative contact with the current experimental findings,
it is crucial to take into account the trapping potential that is
necessary to prevent the atoms from escaping. Therefore, the theoretical
analysis is more complicated. The simplest way to incorporate the
effect of the trap is to use a local density approximation (LDA),
where the system is treated locally as being homogeneous, with spatial
variation included via a local chemical potential that includes the
trap potential. Although this method has been extensively used to
study the density profiles of spin-polarized Fermi gases \cite{chevy,yi,silva,haquea,yip2,imambekov,martikainen,chien2,bulgac2,gubbels},
its validity has never been thoroughly examined.

Alternatively, within the mean-field approximation, one may adopt
the Bogoliubov-de Gennes (BdG) equations, which include the full spatially
varying trap potential from the outset\cite{mizushima,castorina,kinnunen,machida,jensen}.
It has been claimed that the solution of BdG equations includes FFLO
states with spatially varying order parameter \cite{castorina,kinnunen,machida,jensen}.
However, the only evidence for this statement is the observation that
in such solutions the sign and amplitude of the order parameter or
gap function exhibit a small oscillation near the edge of traps.

In this paper, we perform a \emph{comparative} study of the thermodynamic
properties of a trapped, spin-polarized Fermi gas, by using \emph{both}
the mean-field BdG equations and the LDA. While neither method is
exact, since pairing fluctuations beyond the mean-field approximation
are neglected, this comparison can give insight into the consequences
of the different types of approximation in common use. As the most
interesting experiments were in the BCS-BEC crossover regime, we focus
on the on-resonance situation. In this regime - sometimes called the
unitary limit \cite{ho2004}, the $s$-wave scattering length diverges.

The purpose of this work is three-fold. First of all, we have developed
an efficient and accurate hybrid method for solving the mean-field
BdG equations, based on the combined use of a mode expansion in a
finite basis for low-lying states, together with a semiclassical approximation
for the highly excited modes beyond a suitably chosen energy cut-off.
The cut-off is then varied to check the accuracy of the hybrid approach.
As a consequence, we are able to consider a Fermi gas with a large
number of atoms ($\sim10^{5}$) that is of the same order as in experiment.

The second purpose is to use this gain in efficiency to perform a
detailed check on the accuracy of the LDA description in comparison
with the BdG results. It is worth noting that, unlike previous theoretical
work, we do \emph{not} include the Hartree term in the BdG equations,
since this should be {\em unitarity limited} in the BCS-BEC crossover
regime \cite{mit2003}. For a sufficiently large number of atoms,
we find an excellent agreement between these two approximation schemes.
In particular, small oscillations of the order parameter at the edge
of traps which were reported previously, tend to vanish as the number
of particles rises. Therefore, we interpret this as a finite size
effect, rather than the appearance of a spatially modulated FFLO state.

Finally, various thermodynamical quantities are calculated. The observed
bimodal structure in the density distribution of the minority spin
component is reproduced theoretically. The transition temperature
is determined as a function of the spin population imbalance, and
is found to qualitatively match the available experimental data. The
temperature dependence of the equation of state is also discussed.

The paper is organized as follows. In the next section, we present
the theoretical model for a spin-polarized, trapped Fermi gas. In
Sec. III and Sec. IV, we explain the LDA formalism, and then describe
in detail how we solve the BdG equations. The relationship between
these two methods is explored. In Sec. V, a detailed comparison between
LDA and BdG calculations is performed. Results for various thermodynamical
quantities are shown. Their dependence on the population imbalance
and on the temperature are studied. Sec. VI gives our conclusions
and some final remarks.

\section{Models}

We consider a {\em spherically} trapped spin-polarized Fermi gas
at the BCS-BEC crossover point, as found near a molecular Feshbach
resonance. In general, a system like this requires a detailed consideration
of molecule formation channels\cite{Kherunts}, but near a broad Feshbach
resonance, the bound molecular state has a very low population\cite{Hulet1}.
Accordingly, the system can be described approximately by a single
channel Hamiltonian\cite{diener,xiaji}, \begin{eqnarray}
{\cal H} & = & \sum\limits _{\sigma}\int d^{3}{\bf r}\Psi_{\sigma}^{+}\left({\bf r}\right)\left[-\frac{\hbar^{2}}{2m}{\bf \nabla}^{2}+V\left({\bf r}\right)-\mu_{\sigma}\right]\Psi_{\sigma}\left({\bf r}\right)\nonumber \\
 & + & U\int d^{3}{\bf r}\Psi_{\uparrow}^{+}\left({\bf r}\right)\Psi_{\downarrow}^{+}\left({\bf r}\right)\Psi_{\downarrow}\left({\bf r}\right)\Psi_{\uparrow}\left({\bf r}\right),\label{initHami}\end{eqnarray}
 where the pseudospins $\sigma=\uparrow,\downarrow$ denote the two
hyperfine states, and $\Psi_{\sigma}\left({\bf r}\right)$ is the
Fermi field operator that annihilates an atom at position ${\bf r}$
in the spin $\sigma$ state. The number of total atoms is $N=N_{\uparrow}+N_{\downarrow}$.
Two different chemical potentials, $\mu_{\uparrow,\downarrow}=\mu\pm\delta\mu$,
are introduced to take into account the population imbalance $\delta N=N_{\uparrow}-N_{\downarrow}$,
$V\left({\bf r}\right)=m\omega^{2}r^{2}/2$ is the isotropic harmonic
trapping potential with the oscillation frequency $\omega$, and $U$
is the bare inter-atomic interaction strength. We now describe the
LDA and BdG theories.

\section{Local density approximation}

If the number of particles becomes very large, it is natural to assume
that the gas can be divided into many locally uniform sub-systems
with a local chemical potential \cite{yi}. Then, within the LDA,
the trap terms in the Hamiltonian Eq. (\ref{initHami}) are absorbed
into the chemical potential, so that we have effective space-dependent
chemical potentials,

\begin{eqnarray}
\mu_{\uparrow}\left({\bf r}\right) & = & \mu_{\uparrow}-V\left({\bf r}\right),\nonumber \\
\mu_{\downarrow}\left({\bf r}\right) & = & \mu_{\downarrow}-V\left({\bf r}\right).\end{eqnarray}
 Note that the local chemical potential \emph{difference} $\delta\mu\left({\bf r}\right)=\left[\mu_{\uparrow}\left({\bf r}\right)-\mu_{\downarrow}\left({\bf r}\right)\right]/2=\delta\mu$
is always a constant, but the average $\mu\left({\bf r}\right)=\left[\mu_{\uparrow}\left({\bf r}\right)+\mu_{\downarrow}\left({\bf r}\right)\right]/2$
decreases parabolically away from the center of trap.

\subsection{Effective Hamiltonian}

If the global potentials $\mu_{\uparrow}$ and $\mu_{\downarrow}$
are fixed, we can consider a locally uniform Fermi gas in a cell at
position \textbf{r} with local chemical potentials $\mu_{\uparrow}\left({\bf r}\right)$
and $\mu_{\downarrow}\left({\bf r}\right)$, whose Hamiltonian takes
the form, \begin{eqnarray}
{\cal H}\left({\bf r}\right) & = & \sum_{{\bf k}\sigma}\left[\frac{\hbar^{2}k^{2}}{2m}-\mu_{\sigma}\left({\bf r}\right)\right]c_{{\bf k}\sigma}^{+}c_{{\bf k\sigma}}\nonumber \\
 & + & U\sum_{{\bf kk}^{\prime}{\bf p}}c_{{\bf k}\uparrow}^{+}c_{{\bf p}-{\bf k}\downarrow}^{+}c_{{\bf p}-{\bf k}^{\prime}\downarrow}c_{{\bf k}^{\prime}\uparrow}.\label{hamiLDA}\end{eqnarray}
 Here $c_{{\bf k}\sigma}$ represents the annihilation operator for
an atom with kinetic energy $\hbar^{2}k^{2}/(2m)$. For simplicity
we restrict ourself to {\em homogeneous} superfluid states for
the locally uniform cell, \textit{i.e.}, the local order parameter
has zero center-of-mass momentum. By taking the mean-field approximation,
an order parameter of Cooper pairs $\Delta\left({\bf r}\right)=U\sum_{{\bf k}}\left\langle c_{{\bf k}\downarrow}c_{-{\bf k}\uparrow}\right\rangle $
is therefore introduced, whose value depends on position owing to
the spatial dependence of $\mu_{\uparrow}\left({\bf r}\right)$ and
$\mu_{\downarrow}\left({\bf r}\right)$. The local Hamiltonian (\ref{hamiLDA})
then becomes,

\begin{eqnarray}
{\cal H}_{mf}\left({\bf r}\right) & = & \sum_{{\bf k}\sigma}\left[\frac{\hbar^{2}k^{2}}{2m}-\mu_{\sigma}\left({\bf r}\right)\right]c_{{\bf k}\sigma}^{+}c_{{\bf k\sigma}}-\frac{\Delta^{2}\left({\bf r}\right)}{U}\nonumber \\
 & - & \Delta\left({\bf r}\right)\sum_{{\bf k}}\left[c_{{\bf k}\downarrow}c_{-{\bf k}\uparrow}+h.c.\right]\,.\end{eqnarray}
 We have neglected the Hartree terms $U\sum_{{\bf k}}\left\langle c_{{\bf k}\uparrow}^{+}c_{{\bf k}\uparrow}\right\rangle $
and $U\sum_{{\bf k}}\left\langle c_{{\bf k}\downarrow}^{+}c_{{\bf k}\downarrow}\right\rangle $
in this mean-field factorization. Their absence is owing to the following
reasons:

\begin{itemize}
\item The use of contact interactions leads to an unphysical ultraviolet
divergence, and requires a renormalization that expresses the bare
parameter $U$ in terms of the observed or renormalized value $\left(4\pi\hbar^{2}a/m\right)^{-1}$,
\textit{i.e.}, \begin{equation}
\frac{m}{4\pi\hbar^{2}a}=\frac{1}{U}+\sum_{{\bf k}}\frac{1}{2\epsilon_{{\bf k}}},\end{equation}
 where $a$ is the back-ground $s$-wave scattering length between
atoms, and $\epsilon_{{\bf k}}=\hbar^{2}k^{2}/\left(2m\right)$. Generically,
this renormalization requires an infinitely small bare parameter,
in order to compensate the ultraviolet divergence in the summation
$\sum_{{\bf k}}1/2\epsilon_{{\bf k}}$. Therefore, strictly speaking,
within a mean-field approximation the Hartree terms \emph{should}
vanish identically. 
\item For weak couplings, one may indeed obtain Hartree terms like $(4\pi\hbar^{2}a/m)n_{\uparrow,\downarrow}$.
With renormalization, these corrections are beyond mean-field, and
are effective only in the deep BCS limit. Towards the unitarity limit
with increasing scattering length, they are no longer the leading
corrections and become even divergent. Higher order terms are needed
in order to remove the divergence at unitarity. For example, one may
use P\'{a}de approximations in the equation of state\cite{Heiselberg0}. Thus, throughout
the BCS-BEC crossover region, the neglect of the Hartree terms is
not an unreasonable approximation.
\end{itemize}

The above mean-field Hamiltonian can be solved by the standard Bogoliubov
transformation\cite{BdG}. The resulting mean-field thermodynamic
potential has the form,

\begin{eqnarray}
\Omega_{mf}\left({\bf r}\right) & = & -\frac{m}{4\pi\hbar^{2}a}\Delta^{2}\left({\bf r}\right)+\sum_{{\bf k}}\left[\xi_{{\bf k}}-E_{{\bf k}}+\frac{\Delta^{2}\left({\bf r}\right)}{2\epsilon_{{\bf k}}}\right]\nonumber \\
 & + & \frac{1}{\beta}\sum_{{\bf k}}\left[\ln f\left(-E_{{\bf k}+}\right)+\ln f\left(-E_{{\bf k}-}\right)\right],\end{eqnarray}
 where $f\left(x\right)=\left[\exp\left(\beta x\right)+1\right]^{-1}$
is the Fermi distribution function ($\beta=1/k_{B}T$), and $E_{{\bf k}\pm}=E_{{\bf k}}\pm\delta\mu\left({\bf r}\right)$
are the quasiparticle energies with $E_{{\bf k}}=\left[\xi_{{\bf k}}^{2}+\Delta^{2}\left({\bf r}\right)\right]^{1/2}$
and $\xi_{{\bf k}}=\hbar^{2}k^{2}/2m-\mu\left({\bf r}\right)$. Given
local potentials $\mu\left({\bf r}\right)$ and $\delta\mu\left({\bf r}\right)$,
we determine the value of order parameters $\Delta\left({\bf r}\right)$
by minimizing the thermodynamic potential, \textit{i.e.}, $\partial\Omega_{mf}\left({\bf r}\right)/\partial\Delta\left({\bf r}\right)=0,$
or explicitly, \begin{equation}
\frac{m}{4\pi\hbar^{2}a}=\sum_{{\bf k}}\left[\frac{1}{2\epsilon_{{\bf k}}}-\frac{1-f\left(E_{{\bf k}+}\right)-f\left(E_{{\bf k}-}\right)}{2E_{{\bf k}}}\right]\,.\label{gapLDA}\end{equation}
 We note that a non-BCS superfluid solution, the so-called Sarma state
\cite{sarma,wu,pao}, may arise in solving the gap equation (\ref{gapLDA}).
However, on the BCS side such solution suffers from the instabilities
with respect to either the phase separation or a finite-momentum paired
FFLO phase \cite{wu}. Further, the Sarma state is not energetically
favorable \cite{hui2006}, and thereby will be discarded automatically
in the numerical calculations.

\subsection{Thermodynamic Quantities}

Once the local order parameter is fixed, it is straightforward to
calculate the various thermodynamic quantities. The local particle
densities are calculated according to, $n_{\uparrow\text{ }}\left({\bf r}\right)=-\partial\Omega_{mf}\left({\bf r}\right)/\partial\mu_{\uparrow}\left({\bf r}\right)$
and $n_{\downarrow\text{ }}\left({\bf r}\right)=-\partial\Omega_{mf}\left({\bf r}\right)/\partial\mu_{\downarrow}\left({\bf r}\right)$,
or, \begin{eqnarray}
n_{\uparrow}\left({\bf r}\right) & = & \sum_{{\bf k}}\left[\frac{E_{{\bf k}}+\xi_{{\bf k}}}{2E_{{\bf k}}}f\left(E_{{\bf k}-}\right)+\frac{E_{{\bf k}}-\xi_{{\bf k}}}{2E_{{\bf k}}}f\left(-E_{{\bf k}+}\right)\right],\nonumber \\
n_{\downarrow}\left({\bf r}\right) & = & \sum_{{\bf k}}\left[\frac{E_{{\bf k}}+\xi_{{\bf k}}}{2E_{{\bf k}}}f\left(E_{{\bf k}+}\right)+\frac{E_{{\bf k}}-\xi_{{\bf k}}}{2E_{{\bf k}}}f\left(-E_{{\bf k}-}\right)\right],\nonumber \\
 &  & \,\label{nLDA}\end{eqnarray}
 while the entropy and the energy are determined, respectively, by
\begin{eqnarray}
S\left({\bf r}\right) & = & -k_{B}\sum\limits _{{\bf k,}\alpha=\pm}\left[f\left(E_{{\bf k}\alpha}\right)\ln f\left(E_{{\bf k}\alpha}\right)\right.\nonumber \\
 & + & \left.f\left(-E_{{\bf k}\alpha}\right)\ln f_{j}\left(-E_{{\bf k}\alpha}\right)\right],\end{eqnarray}
 and, \begin{equation}
E\left({\bf r}\right)=\Omega_{mf}\left({\bf r}\right)+TS({\bf r})+\mu_{\uparrow}n_{\uparrow}\left({\bf r}\right)+\mu_{\downarrow}n_{\downarrow}\left({\bf r}\right).\end{equation}
 Then, the integration over the whole space gives rise to, \begin{eqnarray}
N\left(\mu,\delta\mu\right) & = & \int d^{3}{\bf r}\left[n_{\uparrow\text{ }}\left({\bf r}\right)+n_{\downarrow\text{ }}\left({\bf r}\right)\right]{\bf ,}\nonumber \\
\delta N\left(\mu,\delta\mu\right) & = & \int d^{3}{\bf r}\left[n_{\uparrow\text{ }}\left({\bf r}\right)-n_{\downarrow\text{ }}\left({\bf r}\right)\right]{\bf ,}\end{eqnarray}
 and $S=\int d^{3}{\bf r}S\left({\bf r}\right)$, $E=\int d^{3}{\bf r}E\left({\bf r}\right)$.
The global chemical potentials $\mu$ and $\delta\mu$ should be adjusted
to satisfy $N\left(\mu,\delta\mu\right)=N$, and $\delta N\left(\mu,\delta\mu\right)=\delta N$.
The numerical calculation thereby involves an iterative procedure.

We note that, on physical grounds, a general picture can be drawn
for the density profiles \cite{silva}. Near the center of the trap,
where $\delta\mu$ is small compared to $\mu\left({\bf r}\right)$,
the densities of the two spin states are forced to be equal and we
have a BCS core extended up to a radius $R_{BCS}$. Outside this radius
a normal state is more favorable than a superfluid phase. At zero
temperature, the Thomas-Fermi radius of the minority (spin down atoms)
and majority (spin up atoms) are given by $R_{TF}^{(1)}=\left[2\left(\mu-\delta\mu\right)/\left(m\omega^{2}\right)\right]^{1/2}$
and $R_{TF}^{(2)}=\left[2\left(\mu+\delta\mu\right)/\left(m\omega^{2}\right)\right]^{1/2}$,
respectively, as we neglect the interactions between the two components
in the normal state.

We finally remark that this entire approach is less accurate at Feshbach
resonance, and especially on the BEC side of resonance, where it becomes
essential to include quantum fluctuations beyond the mean-field approximation\cite{hld}.

\section{The Bogoliubov-de Gennes mean-field theory}

We next consider the Bogoliubov-de Gennes theory of an inhomogeneous
Fermi gas, starting from the Heisenberg equation of motion of Hamiltonian
(\ref{initHami}) for $\Psi_{\uparrow}\left({\bf r},t\right)$ and
$\Psi_{\downarrow}\left({\bf r},t\right)$:

\begin{eqnarray}
i\hbar\frac{\partial\Psi_{\uparrow}}{\partial t} & = & \left[-\frac{\hbar^{2}}{2m}{\bf \nabla}^{2}+V-\mu_{\uparrow}\right]\Psi_{\uparrow}+U\Psi_{\downarrow}^{+}\Psi_{\downarrow}\Psi_{\uparrow},\nonumber \\
i\hbar\frac{\partial\Psi_{\downarrow}}{\partial t} & = & \left[-\frac{\hbar^{2}}{2m}{\bf \nabla}^{2}+V-\mu_{\downarrow}\right]\Psi_{\downarrow}-U\Psi_{\uparrow}^{+}\Psi_{\downarrow}\Psi_{\uparrow}.\nonumber \\
 &  & \,\end{eqnarray}

\subsection{Mean-field approximation}

Within the mean-field approximation as before, we replace the terms
$U\Psi_{\downarrow}^{+}\Psi_{\downarrow}\Psi_{\uparrow}$ and $U\Psi_{\uparrow}^{+}\Psi_{\downarrow}\Psi_{\uparrow}$
by their respective mean-field approximations $U\Psi_{\downarrow}\Psi_{\uparrow}\Psi_{\downarrow}^{+}=-\Delta({\bf r})\Psi_{\downarrow}^{+}+Un_{\downarrow}({\bf r})\Psi_{\uparrow}$
and $U\Psi_{\downarrow}\Psi_{\uparrow}\Psi_{\uparrow}^{+}=-\Delta({\bf r})\Psi_{\uparrow}^{+}+Un_{\uparrow}({\bf r})\Psi_{\downarrow}$,
where we define a gap function $\Delta({\bf r})=-U\langle\Psi_{\downarrow}\Psi_{\uparrow}\rangle$
and $n_{\sigma}({\bf r})=\langle\Psi_{\sigma}^{+}\Psi_{\sigma}\rangle$.
The Hartree terms $Un_{\sigma}({\bf r})$ are infinitely small in
the mean-field treatment due to the regularization of the bare interaction
$U\rightarrow0$, as discussed in greater detail below. We keep them
in the derivation at the moment for clarity. The above decoupling
thus leads to,

\begin{eqnarray}
i\hbar\frac{\partial\Psi_{\uparrow}}{\partial t} & = & \left[{\cal H}_{\uparrow}^{s}-\mu_{\uparrow}\right]\Psi_{\uparrow}-\Delta({\bf r})\Psi_{\downarrow}^{+},\nonumber \\
i\hbar\frac{\partial\Psi_{\downarrow}}{\partial t} & = & \left[{\cal H}_{\uparrow}^{s}-\mu_{\downarrow}\right]\Psi_{\downarrow}+\Delta({\bf r})\Psi_{\uparrow}^{+},\nonumber \\
 &  & \,\end{eqnarray}
 where ${\cal H}_{\sigma}^{s}=-\hbar^{2}{\bf \nabla}^{2}/\left(2m\right)+V\left({\bf r}\right)+Un_{\bar{\sigma}}\left({\bf r}\right)$.
We solve the equation of motion by the insertion of the standard Bogoliubov
transformation:

\begin{eqnarray}
\Psi_{\uparrow} & = & \sum\limits _{j}\left[u_{j\uparrow}\left({\bf r}\right)c_{j\uparrow}e^{-iE_{j\uparrow}t/\hbar/}+v_{j\downarrow}^{*}\left({\bf r}\right)c_{j\downarrow}^{+}e^{iE_{j\downarrow}t/\hbar}\right],\nonumber \\
\Psi_{\downarrow}^{+} & = & \sum\limits _{j}\left[u_{j\downarrow}^{*}\left({\bf r}\right)c_{j\downarrow}^{+}e^{iE_{j\downarrow}t/\hbar}-v_{j\uparrow}\left({\bf r}\right)c_{j\uparrow}e^{-iE_{j\uparrow}t/\hbar}\right]\,.\nonumber \\
 &  & \,\end{eqnarray}
 This yields the well-known BdG equations for the Bogoliubov quasiparticle
wave functions $u_{j\sigma}\left({\bf r}\right)$ and $v_{j\sigma}\left({\bf r}\right)$
with excitation energies $E_{j\sigma}$,

\begin{eqnarray}
\left[\begin{array}{cc}
{\cal H}_{\sigma}^{s}-\mu_{\sigma} & \Delta({\bf r})\\
\Delta^{*}({\bf r}) & -{\cal H}_{\bar{\sigma}}^{s}+\mu_{\bar{\sigma}}\end{array}\right]\left[\begin{array}{c}
u_{j\sigma}\left({\bf r}\right)\\
v_{j\sigma}\left({\bf r}\right)\end{array}\right] & = & E_{j\sigma}\left[\begin{array}{c}
u_{j\sigma}\left({\bf r}\right)\\
v_{j\sigma}\left({\bf r}\right)\end{array}\right],\nonumber \\
 &  & \,\end{eqnarray}
 where $u_{j\sigma}\left({\bf r}\right)$ and $v_{j\sigma}\left({\bf r}\right)$
are normalized by $\int d{\bf r(}\left|u_{j\sigma}\left({\bf r}\right)\right|^{2}+\left|v_{j\sigma}\left({\bf r}\right)\right|^{2})=1$.
The number densities of different hyperfine states $n_{\uparrow}\left({\bf r}\right)=\left\langle \Psi_{\uparrow}^{+}\Psi_{\uparrow}\right\rangle $
and $n_{\downarrow}\left({\bf r}\right)=\left\langle \Psi_{\downarrow}^{+}\Psi_{\downarrow}\right\rangle $,
and the BCS Cooper-pair condensate $\Delta({\bf r})=-U\langle\Psi_{\downarrow}\Psi_{\uparrow}\rangle$,
can then be written as, \begin{eqnarray}
n_{\sigma}\left({\bf r}\right) & = & \frac{1}{2}\sum_{j}\left[\left|u_{j\sigma}\right|^{2}f\left(E_{j\sigma}\right)+\left|v_{j\bar{\sigma}}\right|^{2}f\left(-E_{j\bar{\sigma}}\right)\right],\nonumber \\
\Delta({\bf r}) & = & \frac{U}{2}\sum_{j}\left[v_{j\uparrow}^{*}u_{j\uparrow}f\left(E_{j\uparrow}\right)-u_{j\downarrow}v_{j\downarrow}^{*}f\left(-E_{j\downarrow}\right)\right],\nonumber \\
 &  & \,\label{gap0}\end{eqnarray}
 where the statistical averages $\langle c_{j\sigma}^{+}c_{j\sigma}\rangle=f\left(E_{j\sigma}\right)$
and $\langle c_{j\sigma}c_{j\sigma}^{+}\rangle=f\left(-E_{j\sigma}\right)$
have been used.

The solutions of the BdG equations contain both positive and negative
excitation energies. Thus, to avoid double counting, a factor of half
appears in the summation in Eq. (\ref{gap0}). Furthermore, the presence
of the chemical potential difference breaks the particle-hole symmetry
and therefore leads to different quasiparticle wave functions for
the two components. One can easily identify that there is a one to
one correspondence between the solution for the spin up and spin down
energy levels, \textit{i.e.}, \begin{equation}
E_{j\sigma}\leftrightarrow-E_{j\bar{\sigma}},\end{equation}
 and \begin{equation}
\left[\begin{array}{c}
u_{j\sigma}\left({\bf r}\right)\\
v_{j\sigma}\left({\bf r}\right)\end{array}\right]\leftrightarrow\left[\begin{array}{c}
-v_{j\bar{\sigma}}^{*}\left({\bf r}\right)\\
+u_{j\bar{\sigma}}^{*}\left({\bf r}\right)\end{array}\right].\end{equation}
 By exploiting this symmetry of the BdG equations, therefore, we need
to solve the BdG equations for the spin up part only. This has the
following form after removing the spin index, \textit{i.e.}, we let
$u_{j}\left({\bf r}\right)=u_{j\uparrow}\left({\bf r}\right)$ and
$v_{j}\left({\bf r}\right)=v_{j\uparrow}\left({\bf r}\right)$, to
give: \begin{equation}
\left[\begin{array}{cc}
{\cal H}_{\uparrow}^{s}-\mu_{\uparrow} & \Delta({\bf r})\\
\Delta^{*}({\bf r}) & -{\cal H}_{\downarrow}^{s}+\mu_{\downarrow}\end{array}\right]\left[\begin{array}{c}
u_{j}\left({\bf r}\right)\\
\upsilon_{j}\left({\bf r}\right)\end{array}\right]=E_{j}\left[\begin{array}{c}
u_{j}\left({\bf r}\right)\\
\upsilon_{j}\left({\bf r}\right)\end{array}\right].\label{BdG}\end{equation}

\subsection{Hybrid BdG Technique}

We now wish to address the issue of how to use the BdG equations in
a practical numerical application. Accordingly, the density distributions
and the gap function can be rewritten as, \begin{eqnarray}
n_{\uparrow}\left({\bf r}\right) & = & \sum_{j}\left|u_{j}\left({\bf r}\right)\right|^{2}f\left(E_{j}\right),\nonumber \\
n_{\downarrow}\left({\bf r}\right) & = & \sum_{j}\left|v_{j}\left({\bf r}\right)\right|^{2}f\left(-E_{j}\right),\nonumber \\
\Delta({\bf r}) & = & U\sum_{j}u_{j}\left({\bf r}\right)v_{j}^{*}\left({\bf r}\right)f\left(E_{j}\right).\label{gap}\end{eqnarray}
 Eqs. (\ref{BdG}) and (\ref{gap}) can then be solved self-consistently,
with the constraints that \begin{equation}
N\left(\mu,\delta\mu\right)=\int d^{3}{\bf r}\left[n_{\uparrow\text{ }}\left({\bf r}\right)+n_{\downarrow\text{ }}\left({\bf r}\right)\right]{\bf =}N,\end{equation}
 and \begin{equation}
\delta N\left(\mu,\delta\mu\right)=\int d^{3}{\bf r}\left[n_{\uparrow\text{ }}\left({\bf r}\right)-n_{\downarrow\text{ }}\left({\bf r}\right)\right]{\bf =}\delta N.\end{equation}

In any practical calculation, due to computational limitations, one
has to truncate the summation over the quasiparticle energy levels.
For this purpose, we introduce a hybrid strategy by introducing a
high-energy cut-off $E_{c}$, above which the local density approximation
may be adopted\cite{reidl} for sufficiently high-lying states. Following
this approach, we then have, \begin{eqnarray}
n_{\uparrow}\left({\bf r}\right) & = & n_{\uparrow,d}\left({\bf r}\right)+n_{\uparrow,c}\left({\bf r}\right),\nonumber \\
n_{\downarrow}\left({\bf r}\right) & = & n_{\downarrow,d}\left({\bf r}\right)+n_{\downarrow,c}\left({\bf r}\right),\nonumber \\
\Delta({\bf r}) & = & \Delta_{d}({\bf r})+\Delta_{c}({\bf r}),\end{eqnarray}
 where, \begin{eqnarray}
n_{\uparrow,d}\left({\bf r}\right) & = & \sum_{\left|E_{j}\right|<E_{c}}\left|u_{j}\left({\bf r}\right)\right|^{2}f\left(E_{j}\right),\nonumber \\
n_{\uparrow,c}\left({\bf r}\right) & = & \sum_{\left|E_{j}\right|>E_{c}}\left|u_{j}\left({\bf r}\right)\right|^{2}f\left(E_{j}\right),\nonumber \\
n_{\downarrow,d}\left({\bf r}\right) & = & \sum_{\left|E_{j}\right|<E_{c}}\left|v_{j}\left({\bf r}\right)\right|^{2}f\left(-E_{j}\right),\nonumber \\
n_{\downarrow,c}\left({\bf r}\right) & = & \sum_{\left|E_{j}\right|>E_{c}}\left|v_{j}\left({\bf r}\right)\right|^{2}f\left(-E_{j}\right),\end{eqnarray}
 and \begin{eqnarray}
\Delta_{d}({\bf r}) & = & U\sum_{\left|E_{j}\right|<E_{c}}u_{j}\left({\bf r}\right)v_{j}^{*}\left({\bf r}\right)f\left(E_{j}\right),\nonumber \\
\Delta_{c}({\bf r}) & = & U\sum_{\left|E_{j}\right|>E_{c}}u_{j}\left({\bf r}\right)v_{j}^{*}\left({\bf r}\right)f\left(E_{j}\right).\end{eqnarray}
 We consider below separately the contributions from the quasi-continuous
high-lying states ($\left|E_{j}\right|>E_{c}$) and discrete low-energy
states ($\left|E_{j}\right|<E_{c}$). This allows us to take into
account the spatial variation of the low-lying trapped quasiparticle
wave functions, without having to treat all high-energy states in
this formalism.

\subsection{LDA for high-lying states}

For the BdG equations (\ref{BdG}), the local density approximation
is the leading order of a semiclassical approximation and amounts
to setting

\begin{eqnarray}
u_{j}\left({\bf r}\right) & \rightarrow & u\left({\bf k,r}\right)\exp\left[i{\bf k\cdot r}\right],\nonumber \\
\text{ }v_{j}\left({\bf r}\right) & \rightarrow & v\left({\bf k,r}\right)\exp\left[i{\bf k\cdot r}\right],\nonumber \\
E_{j} & \rightarrow & E\left({\bf k}\right),\end{eqnarray}
 where $u\left({\bf k,r}\right)$ and $v\left({\bf k,r}\right)$ are
normalized by $\left|u\left({\bf k,r}\right)\right|^{2}+\left|v\left({\bf k,r}\right)\right|^{2}=1$,
and the level index {}``$j$'' has now been replaced by a wave vector
${\bf k}$. So the equations in (\ref{BdG}) are reduced to the algebraic
form

\begin{equation}
\left[\begin{array}{cc}
{\cal H}_{\uparrow}^{s}\left({\bf k},{\bf r}\right)-\mu_{\uparrow} & \Delta({\bf r})\\
\Delta^{*}({\bf r}) & -{\cal H}_{\downarrow}^{s}\left({\bf k},{\bf r}\right)+\mu_{\downarrow}\end{array}\right]\left[\begin{array}{c}
u_{{\bf k}}\\
v_{{\bf k}}\end{array}\right]=E\left({\bf k}\right)\left[\begin{array}{c}
u_{{\bf k}}\\
v_{{\bf k}}\end{array}\right],\end{equation}
 where the quasi-classical single particle Hamiltonian is \begin{equation}
{\cal H}_{\sigma}^{s}\left({\bf k},{\bf r}\right)=\hbar^{2}k^{2}/\left(2m\right)+V\left({\bf r}\right)+Un_{\bar{\sigma}}\left({\bf r}\right)\,\,.\end{equation}
 We obtain two branches of the excitation spectra, $E\left({\bf k,+}\right)=E_{{\bf k}}-\delta\mu-U[n_{\uparrow}({\bf r})-n_{\downarrow}({\bf r})]/2\ $,
and $E\left({\bf k,-}\right)=E_{{\bf k}}+\delta\mu+U[n_{\uparrow}({\bf r})-n_{\downarrow}({\bf r})]/2$,
which may be interpreted as the particle and hole contributions, respectively.
Here $E_{{\bf k}}=\left[\xi_{{\bf k}}^{2}+\Delta^{2}({\bf r})\right]^{1/2}$
and $\xi_{{\bf k}}=\hbar^{2}k^{2}/2m+V({\bf r})-\mu+Un({\bf r})/2$.
Note that consistent with the definition in Sec. III, we have reversed
the sign of the excitation spectrum of the hole branch, that is, for
the particle branch $E\left({\bf k}\right)=+E\left({\bf k,+}\right)$,
while for holes $E\left({\bf k}\right)=-E\left({\bf k,-}\right)$.
The eigenfunctions of the two branch solutions are, respectively,

\begin{eqnarray}
u_{{\bf k}}^{2} & = & \frac{1}{2}\left(1+\frac{\xi_{{\bf k}}}{E_{{\bf k}}}\right),\text{ }\nonumber \\
v_{{\bf k}}^{2} & = & \frac{1}{2}\left(1-\frac{\xi_{{\bf k}}}{E_{{\bf k}}}\right),\text{ }\nonumber \\
u_{{\bf k}}v_{{\bf k}}^{*} & = & +\frac{\Delta({\bf r})}{2E_{{\bf k}}},\end{eqnarray}
 and \begin{eqnarray}
u_{{\bf k}}^{2} & = & \frac{1}{2}\left(1-\frac{\xi_{{\bf k}}}{E_{{\bf k}}}\right),\text{ }\nonumber \\
v_{{\bf k}}^{2} & = & \frac{1}{2}\left(1+\frac{\xi_{{\bf k}}}{E_{{\bf k}}}\right),\text{ }\nonumber \\
u_{{\bf k}}v_{{\bf k}}^{*} & = & -\frac{\Delta({\bf r})}{2E_{{\bf k}}},\end{eqnarray}
 Thus, above the energy cut-off, the quasi-continuous contribution
of high-lying states to the density profiles and the gap function
can be obtained by, \begin{eqnarray}
n_{\uparrow,c}\left({\bf r}\right) & = & \sum\limits _{E\left({\bf k,+}\right)>Ec}\frac{f\left[E\left({\bf k,+}\right)\right]}{2}\left(1+\frac{\xi_{{\bf k}}}{E_{{\bf k}}}\right)\nonumber \\
 &  & +\sum\limits _{E\left({\bf k,-}\right)>Ec}\frac{f\left[-E\left({\bf k,-}\right)\right]}{2}\left(1-\frac{\xi_{{\bf k}}}{E_{{\bf k}}}\right),\nonumber \\
n_{\downarrow,c}\left({\bf r}\right) & = & \sum\limits _{E\left({\bf k,+}\right)>Ec}\frac{f\left[-E\left({\bf k,+}\right)\right]}{2}\left(1-\frac{\xi_{{\bf k}}}{E_{{\bf k}}}\right)\nonumber \\
 &  & +\sum\limits _{E\left({\bf k,-}\right)>Ec}\frac{f\left[E\left({\bf k,-}\right)\right]}{2}\left(1+\frac{\xi_{{\bf k}}}{E_{{\bf k}}}\right),\label{ndwcont0}\end{eqnarray}
 and \begin{eqnarray}
\Delta_{c}({\bf r}) & = & U\sum\limits _{E\left({\bf k,+}\right)>Ec}\frac{\Delta({\bf r})}{2E_{{\bf k}}}f\left[E\left({\bf k,+}\right)\right]\nonumber \\
 & - & U\sum\limits _{E\left({\bf k,-}\right)>Ec}\frac{\Delta({\bf r})}{2E_{{\bf k}}}f\left[-E\left({\bf k,-}\right)\right].\label{gapcont0}\end{eqnarray}
 It is worth noting that if we reduce the energy cut-off $E_{c}$
to zero, we recover the LDA expressions for the density profiles in
Sec. III (see, for example, Eq. (\ref{nLDA}) ). Moreover, Eq. (\ref{gapcont0})
reduces to the LDA gap equation (\ref{gapLDA}).

At the other extreme, for a sufficiently \emph{large} energy cut-off
($\beta E_{c}\gg1$), we may discard the Fermi distribution function
in Eqs. (\ref{ndwcont0})-(\ref{gapcont0}). As a result we have the
following simplified gap equations for the above cut-off LDA contributions:
\begin{eqnarray}
n_{\uparrow,c}\left({\bf r}\right) & = & \sum\limits _{E\left({\bf k,-}\right)>Ec}\frac{1}{2}\left(1-\frac{\xi_{{\bf k}}}{E_{{\bf k}}}\right),\nonumber \\
n_{\downarrow,c}\left({\bf r}\right) & = & \sum\limits _{E\left({\bf k,+}\right)>Ec}\frac{1}{2}\left(1-\frac{\xi_{{\bf k}}}{E_{{\bf k}}}\right),\label{ndwcont}\end{eqnarray}
 and \begin{equation}
\Delta_{c}({\bf r})=U\sum\limits _{E\left({\bf k,-}\right)>Ec}\left[-\frac{\Delta({\bf r})}{2E_{{\bf k}}}\right].\label{gapcont}\end{equation}

\subsection{BdG equations for low-lying states}

Let us now turn to the low-lying states by solving the BdG equations
(\ref{BdG}). As we consider a spherical trap, it is convenient to
label the Bogoliubov quasiparticle wave functions $u_{j}\left({\bf r}\right)$
and $v_{j}\left({\bf r}\right)$ in terms of the usual quantum numbers
$j=\{ nlm\}$, and write,

\begin{eqnarray}
u_{j}\left({\bf r}\right) & = & \frac{u_{nl}\left(r\right)}{r}Y_{lm}\left(\theta,\varphi\right),\nonumber \\
\text{ }v_{j}\left({\bf r}\right) & = & \frac{v_{nl}\left(r\right)}{r}Y_{lm}\left(\theta,\varphi\right).\end{eqnarray}
 Here $u_{nl}\left(r\right)/r$ and $v_{nl}\left(r\right)/r$ are
the standard radial wave functions, and $Y_{lm}\left(\theta,\varphi\right)$
is the spherical harmonic function. The BdG equations are then given
by,

\begin{equation}
\left[\begin{array}{cc}
{\cal H}_{\uparrow}^{s}\left(l\right)-\mu_{\uparrow} & \Delta({\bf r})\\
\Delta({\bf r}) & -{\cal H}_{\downarrow}^{s}\left(l\right)+\mu_{\downarrow}\end{array}\right]\left[\begin{array}{c}
u_{nl}\\
v_{nl}\end{array}\right]=E_{nl}\left[\begin{array}{c}
u_{nl}\\
v_{nl}\end{array}\right],\label{harmBdG}\end{equation}
 where \begin{equation}
{\cal H}_{\sigma}^{s}\left(l\right)=\frac{-\hbar^{2}}{2m}\left[\frac{d^{2}}{dr^{2}}+\frac{l\left(l+1\right)}{r^{2}}\right]+V\left(r\right)+Un_{\bar{\sigma}}(r)\end{equation}
 is the single particle Hamiltonian in the $l$ sector. We solve these
equations by expanding $u_{nl}\left(r\right)$ and $v_{nl}\left(r\right)$
with respect to the eigenfunctions $\phi_{\alpha l}\left(r\right)$
of a 3D harmonic oscillator radial Hamiltonian, ${\cal H}_{osc}\left(l\right)=-\hbar^{2}/\left(2m\right)\left[d^{2}/dr^{2}+l\left(l+1\right)/r^{2}\right]+V\left(r\right)$
. These have energy eigenvalues $\epsilon_{\alpha l}=(2\alpha+l+3/2)\hbar\omega$,
and the resulting expansion is: \begin{eqnarray}
u_{nl}\left(r\right) & = & \sum_{\alpha}A_{nl}^{\alpha}\phi_{\alpha l}\left(r\right),\nonumber \\
v_{nl}\left(r\right) & = & \sum_{\alpha}B_{nl}^{\alpha}\phi_{\alpha l}\left(r\right).\end{eqnarray}
 The problem is then converted to obtain the eigenvalues and eigenstate
of a symmetric matrix, \begin{equation}
\left[\begin{array}{cc}
\epsilon_{\alpha l\uparrow}\delta_{\alpha\beta}+M_{\alpha\beta}^{\uparrow} & \Delta_{\alpha\beta}\\
\Delta_{\alpha\beta} & -\epsilon_{\alpha l\downarrow}\delta_{\alpha\beta}-M_{\alpha\beta}^{\downarrow}\end{array}\right]\left[\begin{array}{l}
A_{nl}^{\beta}\\
B_{nl}^{\beta}\end{array}\right]=E_{nl}\left[\begin{array}{l}
A_{nl}^{\alpha}\\
B_{nl}^{\alpha}\end{array}\right],\label{matrix}\end{equation}
 where, we have defined $\epsilon_{\alpha l\sigma}=\epsilon_{\alpha l}-\mu_{\sigma}$,
and: \begin{eqnarray}
\Delta_{\alpha\beta} & = & \int dr\phi_{\alpha l}\left(r\right)\Delta\left({\bf r}\right)\phi_{\beta l}\left(r\right),\nonumber \\
M_{\alpha\beta}^{\uparrow} & = & \int dr\phi_{\alpha l}\left(r\right)Un_{\downarrow}({\bf r})\phi_{\beta l}\left(r\right),\nonumber \\
M_{\alpha\beta}^{\downarrow} & = & \int dr\phi_{\alpha l}\left(r\right)Un_{\uparrow}({\bf r})\phi_{\beta l}\left(r\right).\end{eqnarray}
 We note that the renormalization condition $u_{nl}\left(r\right)$
and $v_{nl}\left(r\right)$, $\int dr\left[u_{nl}^{2}\left(r\right)+v_{nl}^{2}\left(r\right)\right]\equiv1$,
is strictly satisfied, since $\sum_{\alpha}\left(A_{nl}^{\alpha}\right)^{2}+\left(B_{nl}^{\alpha}\right)^{2}=1$.
Once $u_{nl}\left(r\right)$ and $v_{nl}\left(r\right)$ are obtained,
we calculate the gap equation for the low-lying states ($\left|E_{nl}\right|\leq E_{c}$)
and the corresponding number equations:

 \begin{eqnarray}
\Delta_{d}({\bf r}) & = & U\sum\limits _{nl}\frac{2l+1}{4\pi r^{2}}u_{nl}\left(r\right)v_{nl}\left(r\right)f\left(E_{nl}\right)\text{,}\nonumber \\
n_{\uparrow,d}\left({\bf r}\right) & = & \sum\limits _{nl}\frac{2l+1}{4\pi r^{2}}u_{nl}^{2}\left(r\right)f\left(E_{nl}\right),\text{ }\nonumber \\
n_{\downarrow,d}\left({\bf r}\right) & = & \sum\limits _{nl}\frac{2l+1}{4\pi r^{2}}v_{nl}^{2}\left(r\right)f\left(-E_{nl}\right).\label{gapdisc}\end{eqnarray}

\subsection{Regularization of the bare interaction $U$}

We now must replace the bare interaction $U$ by the corresponding
s-wave scattering length, using standard renormalization techniques.
The combination of expressions (\ref{gapcont}) and (\ref{gapdisc})
gives the full gap equation, \begin{equation}
\frac{\Delta({\bf r})}{U}=\sum_{\left|E_{j}\right|<E_{c}}u_{j}\left({\bf r}\right)v_{j}^{*}\left({\bf r}\right)f\left(E_{j}\right)-\sum\limits _{E\left({\bf k,-}\right)>Ec}\frac{\Delta({\bf r})}{2E_{{\bf k}}},\end{equation}
 which is formally ultraviolet divergent due to the use of the contact
potential. However, the form of the second term on the right side
of the equation suggests a simple regularization procedure. We substitute
$1/U=\left(4\pi\hbar^{2}a/m\right)^{-1}-\sum_{{\bf k}}1/2\epsilon_{{\bf k}}$
into the above equation and obtain, \begin{eqnarray}
\frac{m}{4\pi\hbar^{2}a}\Delta({\bf r}) & = & \sum_{\left|E_{j}\right|<E_{c}}u_{j}\left({\bf r}\right)v_{j}^{*}\left({\bf r}\right)f\left(E_{j}\right)\nonumber \\
 & + & \sum_{{\bf k}}\frac{\Delta({\bf r})}{2\epsilon_{{\bf k}}}-\sum\limits _{E\left({\bf k,-}\right)>Ec}\frac{\Delta({\bf r})}{2E_{{\bf k}}}.\end{eqnarray}
 Thus we may rewrite the gap equation in terms of an effective coupling
constant $U_{eff}\left({\bf r}\right)$, \textit{i.e.}, \begin{equation}
\Delta({\bf r})=U_{eff}\left({\bf r}\right)\sum_{\left|E_{j}\right|<E_{c}}u_{j}\left({\bf r}\right)v_{j}^{*}\left({\bf r}\right)f\left(E_{j}\right),\label{gapEq}\end{equation}
 where we have introduced an effective coupling constant $U_{eff}\left({\bf r}\right)$
defined so that: \begin{equation}
\frac{1}{U_{eff}\left({\bf r}\right)}=\frac{m}{4\pi\hbar^{2}a}-\left[\sum_{{\bf k}}\frac{1}{2\epsilon_{{\bf k}}}-\sum\limits _{E\left({\bf k,-}\right)>Ec}\frac{1}{2E_{{\bf k}}}\right].\label{regPrep}\end{equation}
 The ultraviolet divergence now cancels in the bracketed expression.
The effective coupling constant $U_{eff}\left({\bf r}\right)$ will
depend on the cut-off energy. However, the resulting gap in Eq. (\ref{gapEq})
is essentially cut-off {\em independent}.

The use of the uniform regularization relation $1/U=\left(4\pi\hbar^{2}a/m\right)^{-1}-\sum_{{\bf k}}1/2\epsilon_{{\bf k}}$
leads to an infinitely small bare interaction coupling. One therefore
has to replace $U$ by zero anywhere if there is no ultraviolet divergence
in the summations. As mentioned earlier, this replacement is the proper
treatment within mean-field theory. Certainly, this procedure neglects
the Hartree correction, which is of importance in the deep BCS regime.
However, around the unitarity regime of interest here, the usual expression
for the Hartree correction becomes divergent, and requires a more
rigorous theoretical treatment which shows that it is no longer significant\cite{Heiselberg0}.
Consistent with this treatment, we note that these mean-field Hartree
shifts are {\em not} observed experimentally in the energy spectra
in the BCS-BEC crossover regime \cite{mit2003}. In other words, the
Hartree terms should be {\em unitarity limited} at crossover.

It is important to point that in principle, the regularization procedure
proposed above is equivalent to the use of a pseudopotential, as suggested
by Bruun and co-workers \cite{bruun}. However, the pseudopotential
regularization involves a calculation of the regular part of the Green
function associated with the single particle Hamiltonian ${\cal H}_{s}$
and is numerically inefficient. Alternative simplified regularization
procedures have also been introduced by Bulgac and Yu \cite{bulgac3},
and Grasso and Urban \cite{grasso}. Our prescription (\ref{regPrep})
may be regarded as a formal improvement of these regularization procedures.

\subsection{Summary of BdG formalism}

We now summarize the BdG formalism, by converting the summation over
the momentum ${\bf k}$ in the high-lying levels to a continuous integral
of the energy. 

We find that the total spin densities are given by:\begin{eqnarray}
n_{\uparrow}\left({\bf r}\right) & = & \sum_{\left|E_{nl}\right|<E_{c}}\frac{2l+1}{4\pi r^{2}}u_{nl}^{2}\left(r\right)f\left(E_{nl}\right)\nonumber \\
 & + & \int_{E_{c}}^{\infty}d\epsilon n_{\uparrow,c}\left(\epsilon,{\bf r}\right),\nonumber \\
n_{\downarrow}\left({\bf r}\right) & = & \sum_{\left|E_{nl}\right|<E_{c}}\frac{2l+1}{4\pi r^{2}}v_{nl}^{2}\left(r\right)f\left(-E_{nl}\right)\nonumber \\
 & + & \int_{E_{c}}^{\infty}d\epsilon n_{\downarrow,c}\left(\epsilon,{\bf r}\right),\label{numberBdG}\end{eqnarray}
with a modified gap equation of:\begin{eqnarray}
\frac{\Delta({\bf r})}{U_{eff}\left({\bf r}\right)} & = & \sum_{\left|E_{nl}\right|<E_{c}}\frac{2l+1}{4\pi r^{2}}u_{nl}\left(r\right)v_{nl}\left(r\right)f\left(E_{nl}\right).\nonumber \\
 & \,\label{gapBdG}\end{eqnarray}
 The above-cutoff contributions are given by: \begin{eqnarray}
n_{\uparrow,c}\left(\epsilon,{\bf r}\right) & = & \frac{\sqrt{2}m^{3/2}}{4\pi^{2}\hbar^{3}}\left[\frac{\epsilon+\delta\mu}{\sqrt{\left(\epsilon+\delta\mu\right)-\Delta^{2}({\bf r})}}-1\right]\nonumber \\
 & \times & \left[\sqrt{\left(\epsilon+\delta\mu\right)-\Delta^{2}({\bf r})}+\mu-V\right]^{1/2},\nonumber \\
n_{\downarrow,c}\left({\bf r}\right) & = & \frac{\sqrt{2}m^{3/2}}{4\pi^{2}\hbar^{3}}\left[\frac{\epsilon-\delta\mu}{\sqrt{\left(\epsilon-\delta\mu\right)-\Delta^{2}({\bf r})}}-1\right]\nonumber \\
 &\times  & \left[\sqrt{\left(\epsilon-\delta\mu\right)-\Delta^{2}({\bf r})}+\mu-V\right]^{1/2},\nonumber \\
 &  & \,\end{eqnarray}
 and the value under the square root is understood to be non-negative.
Moreover, in an integral form the effective coupling takes the form,
\begin{equation}
\frac{1}{U_{eff}\left({\bf r}\right)}=\frac{m}{4\pi\hbar^{2}a}-\frac{k_{c}}{2\pi^{2}}-\frac{\sqrt{2}m^{3/2}}{4\pi^{2}\hbar^{3}}\int_{E_{c}}^{\infty}d\epsilon f\left(\epsilon,{\bf r}\right),\label{ueff}\end{equation}
 where, \begin{eqnarray}
k_{c} & = & \left[\sqrt{\left(E_{c}-\delta\mu\right)-\Delta^{2}({\bf r})}+\mu-V\right]^{1/2},\nonumber \\
f\left(\epsilon,{\bf r}\right) & = & \frac{\left[\sqrt{\left(\epsilon-\delta\mu\right)-\Delta^{2}({\bf r})}+\mu-V\right]^{1/2}}{\sqrt{\left(\epsilon-\delta\mu\right)-\Delta^{2}({\bf r})}}\nonumber \\
 & \times & \left[\frac{\epsilon-\delta\mu}{\sqrt{\left(\epsilon-\delta\mu\right)-\Delta^{2}({\bf r})}+\mu-V}-1\right].\nonumber \\
 &  & \,\end{eqnarray}
 The radial wavefunctions in Eqs. (\ref{numberBdG}) and (\ref{gapBdG}) are calculated by
solving the eigenvalue problem (\ref{matrix}). As the matrix involves
the gap function, a self-consistent iterative procedure is necessary.
For a given number of atoms ($N=N_{\uparrow}+N_{\downarrow}$ and
$\delta N=N_{\uparrow}-N_{\downarrow}$), temperature and s-wave scattering
length, we:

\begin{enumerate}
\item start with the LDA results or a previously determined better estimate
for $\Delta\left({\bf r}\right)$, 
\item solve Eq. (\ref{ueff}) for the effective coupling constant, 
\item then solve Eq. (\ref{matrix}) for all the radial states up to the
chosen energy cut-off to find $u_{nl}\left(r\right)$ and $v_{nl}\left(r\right)$,
and 
\item finally determine an improved value for the gap function from Eq.
(\ref{gapBdG}). 
\end{enumerate}
During the iteration, the density profiles $n_{\uparrow\text{ }}\left({\bf r}\right)$
and $n_{\downarrow\text{ }}\left({\bf r}\right)$ are updated. The
chemical potentials $\mu$ and $\delta\mu$ are also adjusted slightly
in each iterative step to enforce the number-conservation condition
that $\int_{0}^{\infty}dr4\pi r^{2}\left[n_{\uparrow\text{ }}\left({\bf r}\right)+n_{\downarrow\text{ }}\left({\bf r}\right)\right]{\bf =}N$
and $\int_{0}^{\infty}dr4\pi r^{2}\left[n_{\uparrow\text{ }}\left({\bf r}\right)-n_{\downarrow\text{ }}\left({\bf r}\right)\right]{\bf =}\delta N$,
when final convergence is reached. To make contact with the experimental
observed density profiles \cite{mit2006b,rice}, we calculate the
axial and radial column densities, \begin{eqnarray}
n_{\uparrow}\left(\rho\right) & = & \int_{-\infty}^{\infty}dzn_{\uparrow}\left(\sqrt{\rho^{2}+z^{2}}\right),\nonumber \\
n_{\downarrow}\left(\rho\right) & = & \int_{-\infty}^{\infty}dzn_{\downarrow}\left(\sqrt{\rho^{2}+z^{2}}\right),\end{eqnarray}
 and \begin{eqnarray}
n_{\uparrow}\left(z\right) & = & \int_{0}^{\infty}2\pi\rho d\rho n_{\uparrow}\left(\sqrt{\rho^{2}+z^{2}}\right),\nonumber \\
n_{\downarrow}\left(z\right) & = & \int_{0}^{\infty}2\pi\rho d\rho n_{\downarrow}\left(\sqrt{\rho^{2}+z^{2}}\right).\end{eqnarray}

\subsection{Entropy and energy}

Apart from the density profiles and gap function, we can also determine
the entropy and total energy of the imbalanced Fermi gas, by using
the expressions: \begin{eqnarray}
S & = & -k_{B}\sum_{nl}\left(2l+1\right)\left[f\left(E_{nl}\right)\ln f\left(E_{nl}\right)\right.\nonumber \\
 &  & \left.+f\left(-E_{nl}\right)\ln f\left(-E_{nl}\right)\right],\end{eqnarray}
 and \begin{equation}
E=\int d^{3}{\bf r}\left\{ \sum_{\sigma}\left[\Psi_{\sigma}^{+}\left({\bf r}\right){\cal H}_{\sigma}^{s}\Psi_{\sigma}\left({\bf r}\right)\right]-\frac{\left|\Delta({\bf r})\right|^{2}}{U}\right\} .\end{equation}
 The energy can further be written as, \begin{eqnarray}
E & = & \left[\mu_{\uparrow}N_{\uparrow}+\mu_{\downarrow}N_{\downarrow}-\frac{m}{4\pi\hbar^{2}a}\int d^{3}{\bf r}\left|\Delta\right|^{2}\right]\nonumber \\
 & + & \sum_{j}E_{j}\left[f\left(E_{j}\right)-\int d^{3}{\bf r}\left|v_{j}\right|^{2}\right]\nonumber \\
 & + & \int d^{3}{\bf r}\sum_{{\bf k}}\frac{\left|\Delta\right|^{2}}{2\epsilon_{{\bf k}}},\end{eqnarray}
 where we have replaced the bare interaction $U$ by the s-wave scattering
length. The contribution of the high-energy part to the entropy is
essentially zero.

For the total energy, we must take into account both the low-lying
states and the high-lying states. Therefore, we divide the energy
into two parts $E=E_{d}+E_{c}$, where \begin{eqnarray}
E_{d} & = & \left[\mu_{\uparrow}N_{\uparrow}+\mu_{\downarrow}N_{\downarrow}-\frac{m}{4\pi\hbar^{2}a}\int_{0}^{\infty}dr4\pi r^{2}\left|\Delta\right|^{2}\right]\nonumber \\
 & + & \sum_{\left|E_{nl}\right|<E_{c}}(2l+1)E_{nl}\left[f\left(E_{nl}\right)-\int_{0}^{\infty}drv_{nl}^{2}\left(r\right)\right],\nonumber \\
 &  & \,\end{eqnarray}
 and \begin{eqnarray}
E_{c} & = & \int d^{3}{\bf r}\frac{1}{2}\left[-\sum\limits _{E\left({\bf k,+}\right)>Ec}\left(1-\frac{\xi_{{\bf k}}}{E_{{\bf k}}}\right)E\left({\bf k,+}\right)\right.\nonumber \\
 & - & \left.\sum\limits _{E\left({\bf k,-}\right)>Ec}E\left({\bf k,-}\right)\left(1-\frac{\xi_{{\bf k}}}{E_{{\bf k}}}\right)+\sum_{{\bf k}}\frac{\left|\Delta\right|^{2}}{\epsilon_{{\bf k}}}\right].\nonumber \\
 &  & \,\end{eqnarray}
 By converting the summation into an integral, we obtain $E_{c}=\int_{0}^{\infty}dr4\pi r^{2}E_{c}\left({\bf r}\right)$,
where \begin{eqnarray}
E_{c}\left({\bf r}\right) & = & \frac{\left|\Delta({\bf r})\right|^{2}}{2}\left\{ \frac{k_{c,1}}{2\pi^{2}}+\frac{k_{c,2}}{2\pi^{2}}\right.\nonumber \\
 & - & \left.\int_{E_{c}}^{\infty}d\epsilon\left[E_{c,1}\left(\epsilon,{\bf r}\right)+E_{c,2}\left(\epsilon,{\bf r}\right)\right]\right\} ,\end{eqnarray}

\begin{eqnarray}
k_{c,1} & = & \left[\sqrt{\left(E_{c}+\delta\mu\right)-\Delta^{2}({\bf r})}+\mu-V\left({\bf r}\right)\right]^{1/2},\nonumber \\
k_{c,2} & = & \left[\sqrt{\left(E_{c}-\delta\mu\right)-\Delta^{2}({\bf r})}+\mu-V\left({\bf r}\right)\right]^{1/2},\nonumber \\
 &  & \,\end{eqnarray}
 and \begin{eqnarray}
E_{c,1}\left(\epsilon,{\bf r}\right) & = & \frac{\sqrt{2}m^{3/2}}{2\pi^{2}\hbar^{3}}\frac{\left[E_{+}+\mu-V\left({\bf r}\right)\right]^{1/2}}{E_{+}}\nonumber \\
 &  \times & \left[\frac{\epsilon}{\epsilon+\delta\mu+E_{+}}-\frac{\left(\epsilon+\delta\mu\right)/2}{E_{+}+\mu-V\left({\bf r}\right)}\right],\nonumber \\
E_{c,2}\left(\epsilon,{\bf r}\right) & = & \frac{\sqrt{2}m^{3/2}}{2\pi^{2}\hbar^{3}}\frac{\left[E_{-}+\mu-V\left({\bf r}\right)\right]^{1/2}}{\sqrt{\left(\epsilon-\delta\mu\right)-\Delta^{2}({\bf r})}}\nonumber \\
 &  \times & \left[\frac{\epsilon}{\epsilon-\delta\mu+E_{-}}-\frac{\left(\epsilon-\delta\mu\right)/2}{E_{-}+\mu-V\left({\bf r}\right)}\right],\nonumber \\
 &  & \,\end{eqnarray}
 where $E_{\pm}=\sqrt{\left(\epsilon\pm\delta\mu\right)-\Delta^{2}({\bf r})}$.

\section{Numerical results and discussions}

To be concrete, we will focus on the on-resonance (unitarity) situation
in our numerical calculation, in which the $s$-wave scattering length
goes to infinity. It will be convenient to use {}``trap units'',
\textit{i.e.}, \begin{equation}
m=\omega=\hbar=k_{B}=1.\end{equation}
 Therefore, the length and energy will be measured in units of the
harmonic oscillator length $a_{ho}=\left[\hbar/\left(m\omega\right)\right]^{1/2}$
and $\hbar\omega$, respectively. The temperature is then taken in
units of $\hbar\omega/k_{B}$. It is also illustrative to define some
characteristic scales, considering a spherically trapped {\em ideal}
Fermi gas with equal populations in two hyperfine states (\textit{i.e.},
$N_{\uparrow}=N_{\downarrow}=N/2$). At zero temperature, a simple
LDA treatment of the harmonic potential leads to a Fermi energy $E_{F}=\left(3N\right)^{1/3}\hbar\omega$
and a Fermi temperature $T_{F}=\left(3N\right)^{1/3}\hbar\omega/k_{B}$.
Accordingly, the Thomas-Fermi (TF) radius of the gas is $r_{TF}=\left(24N\right)^{1/6}a_{ho}$,
and the central density for a single species is $n_{TF}(0)=\left(24N\right)^{1/2}/\left(3\pi^{2}\right)a_{ho}^{-3}$.

In a uniform situation, the universality argument gives $\mu=\xi\epsilon_{F}$
in the unitary limit \cite{ho2004}, where $\epsilon_{F}=\hbar^{2}k_{F}^{2}/(2m)$
with a Fermi wavelength $k_{F}=\left(3\pi^{2}n\right)^{1/3}$. The
mean-field BCS theory predicts $\xi\approx0.59$. Therefore, a better
mean-field estimate of the chemical potential and the TF radius for
a trapped unitarity Fermi gas will be $\mu_{TF,unitarity}=\xi^{1/2}E_{F}$
and $r_{TF,unitarity}=\xi^{1/4}r_{TF}$.

For most calculations, we use a number of total atoms $N=20000$,
which is one or two orders of magnitude smaller than in current experiments.
We take a cut-off energy $E_{c}=64\hbar\omega$, which is already
large enough because of the high efficiency of our hybrid strategy.
Typically we solve the BdG equations (\ref{matrix}) within the subspace
$n<n_{\max}=72$ and $l<l_{\max}=120$. The value of $n_{\max}$ and
$l_{\max}$ is determined in such a way that the subspace contains
all the energy levels below $E_{c}$. The calculations usually take
a few hours for a single run in a single-core computer with 3.0GHz
CPU, for a given set of parameters $T$, $N$, $\delta N$, and $4\pi\hbar^{2}a/m$.
Further increase of the number of total atoms to $10^{5}$ or $10^{6}$
is possible, but very time-consuming.

Below we will present some numerical results. In particular we will
examine the validity of the LDA at zero temperature and at finite
temperature. We analyse some previous suggestions of the apparent
appearance of the FFLO states in the mean-field BdG theory. Finally,
we will investigate the thermodynamical behavior of the spin-polarized
Fermi gas.

%%%%%%%%%%%%%%%%%%%%%%%%%%%%%%%%%%%%%%%%%%%%%%%%%%%%%%%%%%%%%%%%%%%%%%%%%%%%%%%%%
%
\begin{figure*}
\begin{centering}\includegraphics[clip,width=15cm]{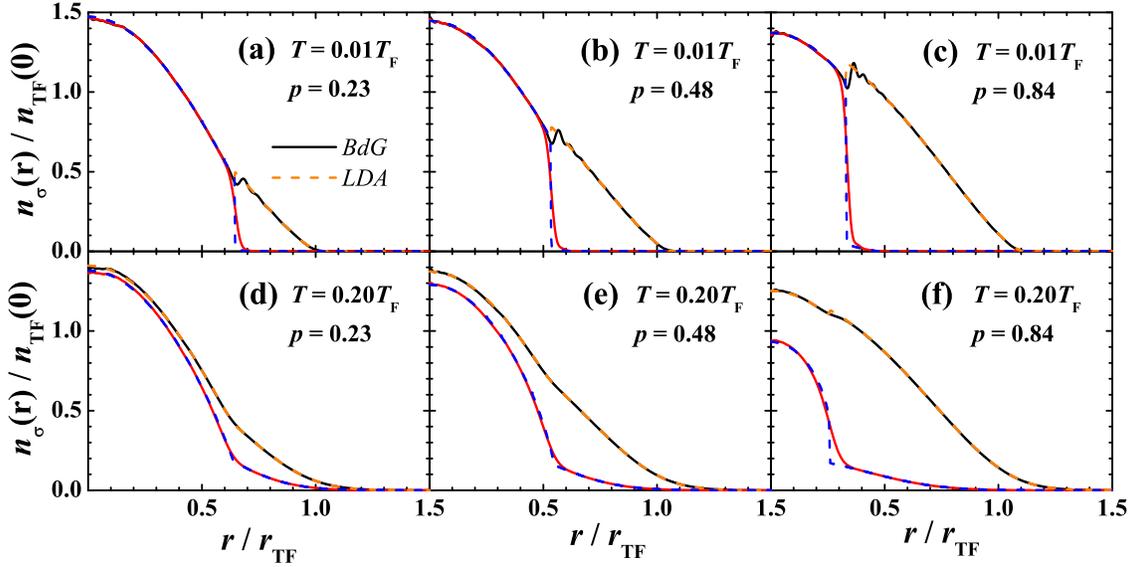}\par\end{centering}

\caption{(color online). Density profiles of spin up and spin down atoms for
a spin-polarized unitary Fermi gas at different population imbalances
and temperatures, as indicated in the figures. The number of total
atoms is $N=20000$. The density profiles are normalized by the Thomas-Fermi
center density of an ideal symmetric Fermi gas with the same number
of total atoms $n_{TF}(0)=\left(24N\right)^{1/2}a_{ho}^{-3}/\left(3\pi^{2}\right)$,
while the length is renormalized by the corresponding Thomas-Fermi
radius $r_{TF}=\left(24N\right)^{1/6}a_{ho}$. The solid lines and
dashed lines refer to the BdG results and LDA results, respectively.}

\label{fig1} 
\end{figure*}

%%%%%%%%%%%%%%%%%%%%%%%%%%%%%%%%%%%%%%%%%%%%%%%%%%%%%%%%%%%%%%%%%%%%%%%%%%%%%%%%%

\subsection{LDA versus BdG}

We present in Fig. 1 the density profiles of two spin states at temperatures
$T=0.01T_{F\text{ }}$ and $T=0.20T_{F\text{ }}$ for different population
imbalances $P=\left(N_{\uparrow}-N_{\downarrow}\right)/\left(N_{\uparrow}+N_{\downarrow}\right)=0.23$,
$0.48$ and $0.84$ as indicated. The results from the BdG and LDA
approaches are plotted using solid lines and dashed lines, respectively.
There is an apparent phase separation phenomenon, with a superfluid
inner core and normal shell outside, which is in consistent with the
recent experimental observation by Zwierlein \textit{et al}. \cite{mit2006b}
and Partridge \textit{et al.} \cite{rice}. Particularly, for $P=0.84$
at $T=0.20T_{F\text{ }}$ the minority (spin down) profile is enhanced
at center, which in turn induces a slight decrease of the central
density of the majority component. The appearance of a dense central
feature in the minority spin profile agrees well with the on-resonance
measurement reported by Zwierlein \textit{et al}. \cite{mit2006b}.
It clearly resembles the bimodal structure in the density distribution
of a BEC.

For all the spin polarizations considered, we find reasonable agreement
between these two methods at the chosen total number of atoms $N=20000$.
As we shall see below, the agreement persists in various thermodynamical
quantities, such as the chemical potential, entropy and total energy.
>From the density profiles, the agreement is excellent at a small or
intermediate population imbalance. The difference between the BdG
and LDA prediction tends to be smaller as the temperature increases.
For a large population imbalance, however, the agreement becomes worse.
This can be understood from the corresponding gap functions as given
in Fig. 2. The gap function in LDA experiences a sudden decrease at
the superfluid-normal interface. With increasing population imbalance,
the drop is much more apparent, and accordingly, the BdG gap function
show a very pronounced oscillation behavior. Therefore, the deviation
of the BdG density profiles from the LDA predictions becomes larger.

%%%%%%%%%%%%%%%%%%%%%%%%%%%%%%%%%%%%%%%%%%%%%%%%%%%%%%%%%%%%%%%%%%%%%%%%%%%%%%%%%
%
\begin{figure}
\begin{centering}\includegraphics[clip,width=7cm]{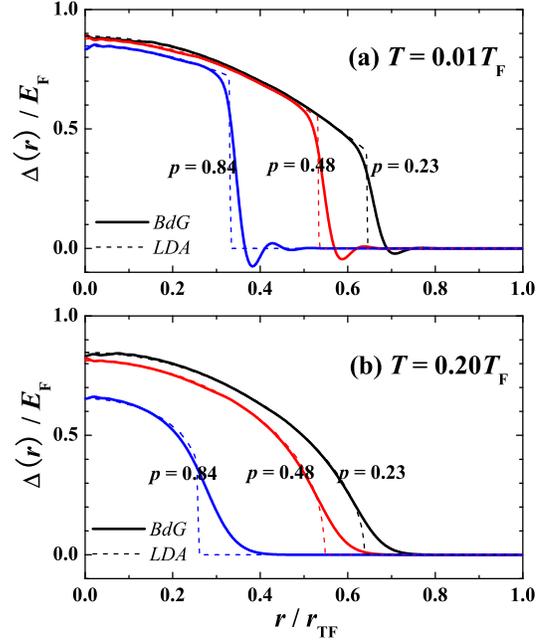}\par\end{centering}

\caption{(color online). Gap functions at different temperatures $T=0.01T_{F}$
(a) and $T=0.20T_{F}$ (b) for various population imbalances as labeled.
The number of total atoms is $N=20000$. The solid lines are the BdG
predictions, while the dashed lines are the LDA results. The value
of gap is renormalized by the non-interacting Fermi energy of an ideal
symmetric Fermi gas $E_{F}=\left(3N\right)^{1/3}\hbar\omega$. Note
that the small oscillation in the gap function at the superfluid-normal
interface at low temperature $T=0.01T_{F}$ vanish when the temperature
becomes high enough.}

\label{fig2} 
\end{figure}

%%%%%%%%%%%%%%%%%%%%%%%%%%%%%%%%%%%%%%%%%%%%%%%%%%%%%%%%%%%%%%%%%%%%%%%%%%%%%%%%%

%%%%%%%%%%%%%%%%%%%%%%%%%%%%%%%%%%%%%%%%%%%%%%%%%%%%%%%%%%%%%%%%%%%%%%%%%%%%%%%%%
%
\begin{figure}
\begin{centering}\includegraphics[clip,width=7cm]{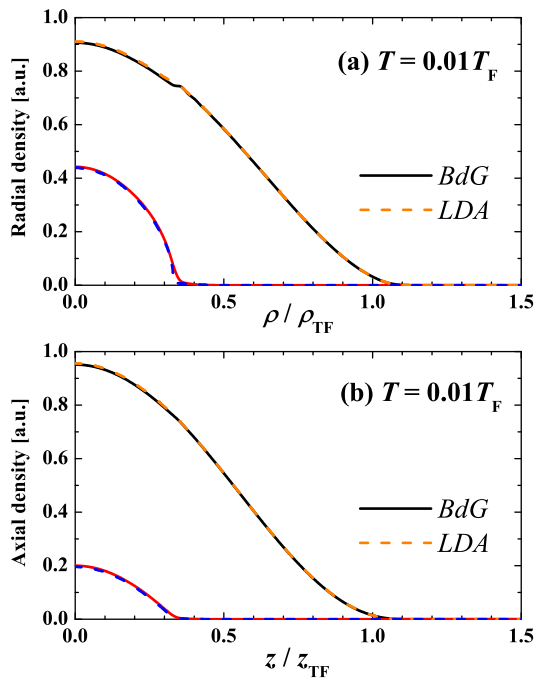}\par\end{centering}

\caption{(color online). Radial (a) and axial (b) column density profiles
for $N=20000$ and $P=0.84$ at $T=0.01T_{F}$. The difference between
the BdG results (solid lines) and LDA results (dashed lines) becomes
extremely small due to the integration over the extra dimensions.}

\label{fig3} 
\end{figure}

%%%%%%%%%%%%%%%%%%%%%%%%%%%%%%%%%%%%%%%%%%%%%%%%%%%%%%%%%%%%%%%%%%%%%%%%%%%%%%%%%

It is important to note that only the axial column density or radial
column density can be measured by the absorption imaging technique
in the experiment. In Figs. 3a and 3b, we plot respectively the axial
column profile and radial column profile at $T=0.01T_{F}$ for the
imbalance $P=0.84$. The minor difference between BdG and LDA shown
in the three-dimensional density profiles is essentially washed out.

%%%%%%%%%%%%%%%%%%%%%%%%%%%%%%%%%%%%%%%%%%%%%%%%%%%%%%%%%%%%%%%%%%%%%%%%%%%%%%%%%
%
\begin{figure}
\begin{centering}\includegraphics[clip,width=8cm]{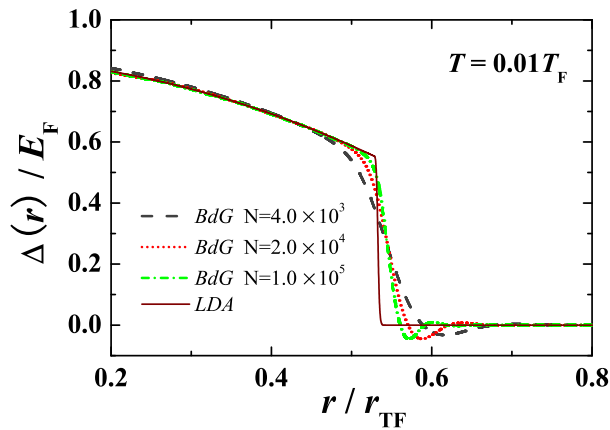}\par\end{centering}

\caption{(color online). Dependence of the gap function on the number of total
atoms at $T=0.01T_{F}$ and $P=0.48$. The solid line is the LDA result,
and the others are BdG results with $N=10^{5}$ (dash-dotted line),
$N=20000$ (dotted line), and $N=4000$ (dashed line).}

\label{fig4} 
\end{figure}

%%%%%%%%%%%%%%%%%%%%%%%%%%%%%%%%%%%%%%%%%%%%%%%%%%%%%%%%%%%%%%%%%%%%%%%%%%%%%%%%%

\subsection{FFLO state at the superfluid-normal interface?}

The consistency between BdG and LDA treatments reported here is in
{\em sharp contrast} with some previous studies \cite{castorina,kinnunen,machida,jensen},
where a notable discrepancy of BdG and LDA results is found. In those
studies the small oscillation of the gap functions at the superfluid-normal
interface is interpreted as the appearance of a spatially modulated
FFLO state. Thus, the discrepancy is explained as due to the breakdown
of LDA approximation for FFLO states. From our results, the oscillation
in the order parameters at the interface appears to be a finite size
effect. This idea is supported by the observation that the BdG formalism
naturally reduces to that of LDA if we set the cut-off energy $E_{c}$
to zero as we mentioned earlier. On the other hand, as shown in Fig.
4, if we increase the number of total particles, the oscillation behavior
of gap functions becomes gradually weaker. We thus infer that the
oscillation will vanish finally in the limit of sufficient large number
of atoms.

To understand the discrepancy of BdG and LDA approaches found in previous
studies \cite{castorina,kinnunen,machida,jensen}, several remarks
may be in order. First, in these studies, the mean-field Hartree terms,
\textit{i.e.}, $Un_{\downarrow}\left({\bf r}\right)$ and $Un_{\uparrow}\left({\bf r}\right)$,
appear in the decoupling of the interaction Hamiltonian in the BdG
theory. However, the Hartree terms is absent in the corresponding
LDA treatment. These terms cannot survive in the regularization procedure
of the bare interaction $U$, but are incorrectly included in Refs.
\cite{castorina,kinnunen,machida,jensen} as $\left(4\pi\hbar^{2}a/m\right)n_{\downarrow}\left({\bf r}\right)$
and $\left(4\pi\hbar^{2}a/m\right)n_{\uparrow}\left({\bf r}\right)$,
respectively.

We note that the absence of the mean-field shift due to Hartree terms
in the strongly interacting BCS-BEC crossover regime is already unambiguously
demonstrated experimental by Gupta \textit{et al.} \cite{mit2003}.
We conclude that there are three main reasons for the differences
between the conclusion we find here that the two approaches are compatible,
as opposed to earlier conclusions to the contrary:

\begin{enumerate}
\item The incorrect inclusion of Hartree terms in one approach but not in
the other, is the most likely reason for the discrepancy of BdG and
LDA results shown in these previous works. 
\item The accuracy of numerical results depends crucially on the regularization
procedure used to treat the bare interaction $U$. Proper treatment
of regularization is essential. 
\item For a large population imbalance, the BdG equations converge very
slowly. Thus, a rigorous criterion is required to ensure complete
convergence. 
\end{enumerate}
%%%%%%%%%%%%%%%%%%%%%%%%%%%%%%%%%%%%%%%%%%%%%%%%%%%%%%%%%%%%%%%%%%%%%%%%%%%%%%%%%
%
\begin{figure}
\begin{centering}\includegraphics[clip,width=7cm]{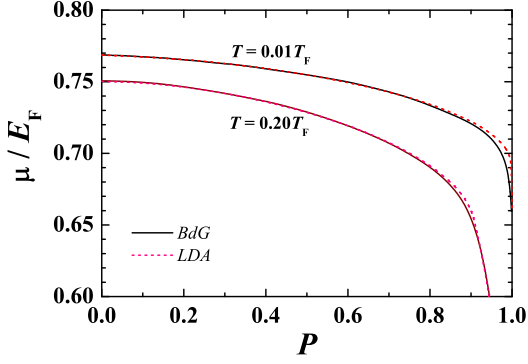}\par\end{centering}

\caption{(color online). Chemical potentials as a function of the population
imbalance at two temperatures $T=0.01T_{F}$ and $T=0.20T_{F}$.}

\label{fig5} 
\end{figure}

%%%%%%%%%%%%%%%%%%%%%%%%%%%%%%%%%%%%%%%%%%%%%%%%%%%%%%%%%%%%%%%%%%%%%%%%%%%%%%%%%

%%%%%%%%%%%%%%%%%%%%%%%%%%%%%%%%%%%%%%%%%%%%%%%%%%%%%%%%%%%%%%%%%%%%%%%%%%%%%%%%%
%
\begin{figure}
\begin{centering}\includegraphics[clip,width=7cm]{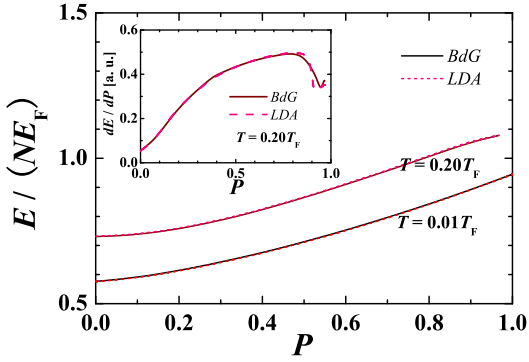}\par\end{centering}

\caption{(color online). Total energy per particle as a function of the population
imbalance $P$ at two temperatures $T=0.01T_{F}$ and $T=0.20T_{F}$.
Inset shows the first order derivation with respect to the population
imbalance at $T=0.20T_{F}$. The jump at $P\sim0.9$ marks the phase
transition to the normal state.}

\label{fig6} 
\end{figure}

%%%%%%%%%%%%%%%%%%%%%%%%%%%%%%%%%%%%%%%%%%%%%%%%%%%%%%%%%%%%%%%%%%%%%%%%%%%%%%%%%

%%%%%%%%%%%%%%%%%%%%%%%%%%%%%%%%%%%%%%%%%%%%%%%%%%%%%%%%%%%%%%%%%%%%%%%%%%%%%%%%%
%
\begin{figure}
\begin{centering}\includegraphics[clip,width=7cm]{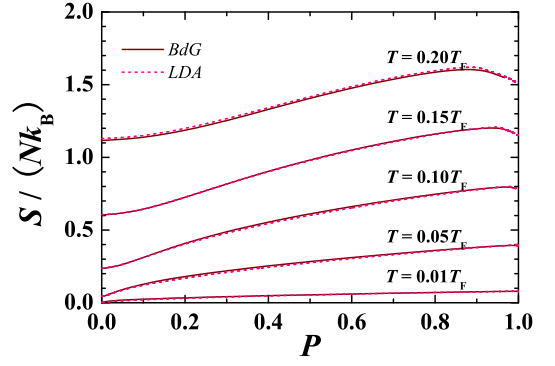}\par\end{centering}

\caption{(color online). Entropy per particle as a function of the population
imbalance at various temperatures.}

\label{fig7} 
\end{figure}

%%%%%%%%%%%%%%%%%%%%%%%%%%%%%%%%%%%%%%%%%%%%%%%%%%%%%%%%%%%%%%%%%%%%%%%%%%%%%%%%%

\subsection{Thermodynamic behavior of the imbalanced Fermi gas at unitarity}

We finally discuss the thermodynamics of the gas at unitarity. In
Figs. 5, 6, and 7, we graph the chemical potential, total energy and
entropy of the gas as a function of the population imbalance at various
temperatures. Again, we find good agreement between the BdG results
and LDA predictions. With increasing population imbalance, the chemical
potential decreases. For a given temperature there is a critical imbalance,
beyond which the Fermi gas transforms into a fully normal state. The
decrease of chemical potential becomes very significant when the phase
transition occurs. In contrast, as population imbalance increases,
the total energy increases. The impact of the phase transition on
the total energy can scarcely be identified since the transition is
smooth.

As shown in the inset of Fig. 6, it \emph{can} be exhibited clearly
in the first order derivative of the energy with respect to the population
imbalance, resembling the behavior of the specific heat as a function
of temperature. The entropy, on the other hand, shows a non-monotonic
dependence at a temperature $\sim0.20T_{F}$. We identify this peak
position as the phase transition point.

%%%%%%%%%%%%%%%%%%%%%%%%%%%%%%%%%%%%%%%%%%%%%%%%%%%%%%%%%%%%%%%%%%%%%%%%%%%%%%%%%
%
\begin{figure}
\begin{centering}\includegraphics[clip,width=7cm]{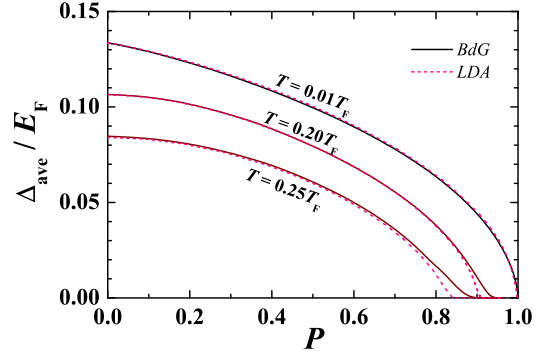}\par\end{centering}

\caption{(color online). Average order parameter defined in Eq. (\ref{averagegap})
as a function of the population imbalance at three temperatures $T=0.01T_{F}$,
$T=0.20T_{F}$, and $T=0.25T_{F}$. The position where the average
order parameter vanishes determines the critical population imbalance
towards the normal state at a fixed temperature.}

\label{fig8} 
\end{figure}

%%%%%%%%%%%%%%%%%%%%%%%%%%%%%%%%%%%%%%%%%%%%%%%%%%%%%%%%%%%%%%%%%%%%%%%%%%%%%%%%%

%%%%%%%%%%%%%%%%%%%%%%%%%%%%%%%%%%%%%%%%%%%%%%%%%%%%%%%%%%%%%%%%%%%%%%%%%%%%%%%%%
%
\begin{figure}
\begin{centering}\includegraphics[clip,width=7cm]{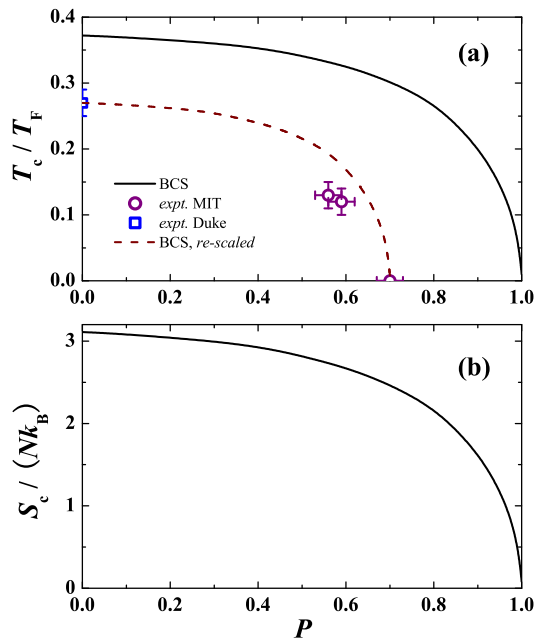}\par\end{centering}

\caption{(color online). (a) BCS mean-field critical temperature as a function
the population imbalance, determined from the LDA calculations (solid
line). Open symbols are available experimental points from Zwierlein \textit{et al.}
\cite{mit2006b,mit2006c} and Kinast \textit{et al.} \cite{duke2005}.
The dashed line shows a re-scaled BCS critical temperature as described
in the text. (b) The critical entropy (the entropy at the critical
temperature) against the population imbalance.}

\label{fig9} 
\end{figure}

%%%%%%%%%%%%%%%%%%%%%%%%%%%%%%%%%%%%%%%%%%%%%%%%%%%%%%%%%%%%%%%%%%%%%%%%%%%%%%%%%

To determine accurately the critical population imbalance as a function
of temperature or vice versa, we plot in Fig. 8 an averaged order
parameter at different temperature, which is defined via,

\begin{equation}
\Delta_{ave}=\left[\frac{\int_{0}^{\infty}dr4\pi r^{2}\Delta^{2}(r)}{N}\right]^{1/2}.\label{averagegap}\end{equation}
 The condition $\Delta_{ave}=0$ therefore determines the critical
population imbalance $P_{c}$ at a fixed temperature. The resulting
$P_{c}$ from the BdG calculation depends, of course, on the number
of total atoms. To remove this dependence, we present the LDA prediction
for critical temperature in Fig. 9a. To compare with the experiments,
we also show four known experimental points in the figure. The discrepancy
between our theoretical predictions and the experimental data is most
likely attributed to the strong pair fluctuations beyond the mean-field,
\textit{i.e.}, for a symmetric gas, the fluctuation shifts $T_{c,BCS}$
from $0.37T_{F}$ to around $0.27T_{F}$ \cite{duke2005}, while at
zero temperature, it reduce $P_{c,BCS}$ from $1.0$ to about $0.70$
\cite{mit2006a,mit2006b}.

Pair fluctuations \emph{must} be taken into account in the strongly
interacting unitarity regime for quantitative comparisons, as we have
shown in earlier calculations \cite{hld} without spin-polarization.
This is certainly the most challenging problem in BCS-BEC crossover
physics. Naively, we may linearly rescale the BCS critical temperature
and population imbalance in such a way that both the theoretical $T_{c}$
at $P=0$ and $P_{c}$ at $T=0$ fits the experimental data. The third
and fourth experimental points of critical temperature at $P=0.56(3)$
and $P=0.59(3)$ now agree approximately with the re-scaled BCS $T_{c}$
curve.

For strongly interacting Fermi gases at low temperature, there is
generally no reliable thermometry technique. The bimodal structure
in the density profile of the minority component may provide a useful
method to determine temperature \cite{mit2006b}. However its accuracy
requires further theoretical investigations due to the strongly correlated
nature of the gas. Entropy is an alternative quantity that may be
used to characterize the temperature in adiabatic passage experiments \cite{hldT}.
In Fig. 9b, we show also the critical entropy of a trapped unitary
gas against the population imbalance. The calculated dependence is
essentially similar to that for temperature.

\section{Conclusions}

We have developed an efficient and accurate hybrid procedure to solve
the mean-field BdG equations for a spherically trapped Fermi gas with
spin population imbalance. This enables us to thoroughly examine the
extensively used LDA approach. For a moderate large particle number
($\sim10^{4}$), the LDA appears to work very well. The discrepancy
of BdG and LDA results reported in previous studies is attributed
to the incorrect inclusion of a mean-field Hartree term in the BdG
equations. We note, however, that the trap used in the current experiments
is elongated in the axial direction. The spherical trap considered
in this work may be regarded as approximately representative of the
experimental setup by Zwierlein \textit{et al.}~\cite{mit2006a,mit2006b,mit2006c} 
with an aspect ratio approximately 5. The trap in the experiment by Partridge 
\textit{et al.}~\cite{rice} is extremely anisotropic, and therefore 
the LDA description could break down. The solution of the BdG equations 
for such elongated systems is numerically intensive, and requires 
further investigation.

Our derivation of the BdG formalism and the numerical results with
varying particle number in Fig. 4 strongly suggest that the calculated
small oscillation in the order parameter at the superfluid-normal
interface arises from finite size effects. This is in marked contrast
with the previous interpretation of this effect as due to a finite-momentum
paired FFLO state \cite{castorina,kinnunen,machida,jensen}. The detailed
structure of the proposed FFLO state is not clear, even in the homogeneous
situation. How to extend the current BdG formalism to incorporate
the FFLO state is therefore a fascinating issue. Another possibility
is that the extremely narrow window for FFLO states in 3D is closed
or reduced in size due to the presence of the harmonic trap.

We have reported various mean-field thermodynamical properties of
the imbalanced Fermi gas at unitarity, which is believed to be qualitatively
reliable at low temperature. The experimentally observed bimodal distribution
in the profile of the minority spin state has been reproduced. We
have determined the BCS superfluid transition temperature as a function
of the population imbalance, and have shown that it is consistent
with recent experimental measurements. Quantitative theories of the
imbalanced Fermi gas that take account of large quantum fluctuations
that occur in the strongly correlated unitarity regime still need
to be developed. A promising approach is to take into account pair
fluctuations within the ladder approximation \cite{hld}. This problem
will be addressed in a future publication.

%===============================================================

\begin{acknowledgments}
This work is supported by the Australian Research Council Center of
Excellence and by the National Science Foundation of China Grant
No. NSFC-10574080 and the National Fundamental Research Program under 
Grant No. 2006CB921404. 
\end{acknowledgments}
%===============================================================

\end{document}